\documentclass[sigconf]{acmart}

\usepackage{acmart-taps}
\AtBeginDocument{%
  }

\copyrightyear{2025}
\acmYear{2025}
\setcopyright{rightsretained}
\acmConference[CHI EA '25]{Extended Abstracts of the CHI Conference on Human Factors in Computing Systems}{April 26-May 1, 2025}{Yokohama, Japan}
\acmBooktitle{Extended Abstracts of the CHI Conference on Human Factors in Computing Systems (CHI EA '25), April 26-May 1, 2025, Yokohama, Japan}
\acmDOI{10.1145/3706599.3706664}
\acmISBN{979-8-4007-1395-8/25/04}
\acmSubmissionID{5901}

\usepackage{CJKutf8}
\usepackage{enumitem}
\usepackage{url}
\usepackage{comment} 

\begin{document}

\title{PaperWave: Listening to Research Papers as Conversational Podcasts Scripted by LLM}

\author{Yuchi Yahagi}
\email{yahagi@hc.ic.i.u-tokyo.ac.jp}
\orcid{0000-0002-9810-5999}
\affiliation{%
  \institution{The University of Tokyo}
  \city{Bunkyo}
  \state{Tokyo}
  \country{Japan}
}
\affiliation{
  \institution{Doctoral Course Research Fellow of Japan Society for the Promotion of Science}
  \city{Chiyoda}
  \state{Tokyo}
  \country{Japan}
}

\author{Rintaro Chujo}
\email{chujo@hc.ic.i.u-tokyo.ac.jp}
\orcid{0000-0002-4499-7047}
\affiliation{%
  \institution{The University of Tokyo}
  \city{Bunkyo}
  \state{Tokyo}
  \country{Japan}
}

\author{Yuga Harada}
\email{yugaharada@gmail.com}
\orcid{0009-0003-6139-2781}
\affiliation{%
  \institution{Educe Technologies}
  \city{Tokyo}
  \state{}
  \country{Japan}
}

\author{Changyo Han}
\email{hanc@hc.ic.i.u-tokyo.ac.jp}
\orcid{0000-0002-9925-3010}
\affiliation{%
  \institution{The University of Tokyo}
  \city{Bunkyo}
  \state{Tokyo}
  \country{Japan}
}

\author{Kohei Sugiyama}
\email{sugiyama_kohei@iii.u-tokyo.ac.jp}
\orcid{0009-0003-3038-5466}
\affiliation{%
  \institution{The University of Tokyo}
  \city{Bunkyo}
  \state{Tokyo}
  \country{Japan}
}

\author{Takeshi Naemura}
\email{naemura@hc.ic.i.u-tokyo.ac.jp}
\orcid{0000-0002-6653-000X}
\affiliation{%
  \institution{The University of Tokyo}
  \city{Bunkyo}
  \state{Tokyo}
  \country{Japan}
}

\renewcommand{\shortauthors}{Yahagi, Chujo, Harada, Han, Sugiyama, and Naemura}

\begin{abstract}
Listening to audio content, such as podcasts and audiobooks, is one way for people to engage with knowledge.
Listening affords people more mobility than reading by seeing, thereby broadening their learning opportunities.
This study explores the potential applications of large language models (LLMs) to adapt text documents to audio content and addresses the lack of listening-friendly materials for niche content, such as research papers.
LLMs can generate scripts of audio content in various styles tailored to specific needs, such as full-content duration or speech types (monologue or dialogue).
To explore this potential, we developed PaperWave as a prototype that transforms academic paper PDFs into conversational podcasts. 
Our two-month investigation, involving 11 participants (including the authors), employed an autobiographical design, a field study, and a design workshop.
The findings highlight the importance of considering listener interaction with their environment when designing document-to-audio systems.
\end{abstract}


\begin{teaserfigure}
\centering
  \includegraphics[width=\textwidth]{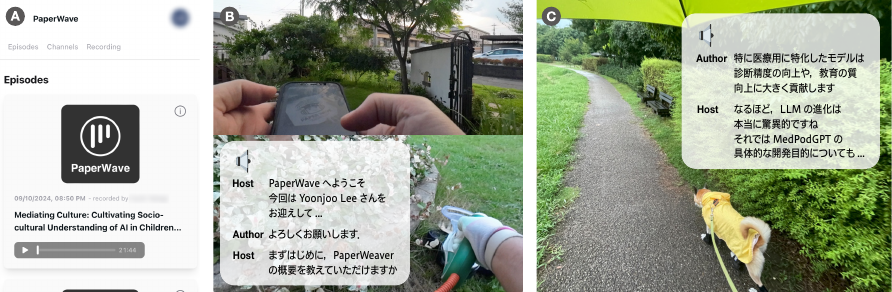}
  \caption{PaperWave application and its real-world usage. 
   (A) Displays the PaperWave web application on a smartphone, which converts research paper PDFs into conversational podcasts. 
   Users can select and listen to generated episodes on this interface. 
   (B) Shows the first author (A1) listening to a podcast in Japanese, A1's first language, while mowing the lawn on August 18. 
   The podcast featured PaperWeaver by Lee et al.~\cite{Lee.2024.PaperWeaverEnrichingTopicalPaper}, originally written in English. 
   An English translation of the script includes: 
   Host: ``Welcome to PaperWave, this time welcoming Yoonjoo Lee...'' 
   Author: ``Nice to meet you.'' 
   Host: ``First of all, can you give us an overview of PaperWeaver?'' 
   (C) Depicts A1 listening to an English paper titled MedPodGPT by Jia et al.~\cite{Jia.2024.MedPodGPTMultilingualAudioaugmentedLarge} in Japanese while walking his dog on August 29. 
   The script is as follows. 
   Author: ``Especially the medical-specific models will contribute greatly to the improvement of diagnostic accuracy and the quality of education.'' 
   Host: ``Well, the evolution of LLM is truly astounding. Now let's talk about the specific development objectives of MedPodGPT...''}
  \Description{Image A: Screenshot of the PaperWave app interface on a mobile device, showing the “Episodes” page. The page displays a list of recorded podcasts with titles and playback controls. The highlighted episode is titled “Mediating Culture: Cultivating Socio-cultural Understanding of AI in Children,” recorded on 09/10/2024 at 08:50 PM, with a duration of 21 minutes and 44 seconds.
Image B: Two first-person images are shown. The top image shows a person holding a smartphone in an outdoor setting, viewing a podcast app. The bottom image depicts a person mowing the lawn with a clipper. The speech bubble overlay shows a conversation in Japanese between a “Host” and an “Author.” The Host welcomes the Author to PaperWave and asks them to explain the overview of PaperWave, to which the Author responds with a greeting.
Image C: A pathway surrounded by greenery, viewed from the perspective of a person walking a dog under a green umbrella. A speech bubble overlay shows another conversation in Japanese. The Author discusses the contribution of specialized models to improving diagnostic accuracy and education in the medical field, while the Host expresses amazement at the advancements of LLMs and asks about the specific development goals of MedPodGPT.}
  \label{fig:teaser}
\end{teaserfigure}


\begin{CJK*}{UTF8}{ipxm}

\maketitle
\makeatletter \gdef\@ACM@checkaffil{} \makeatother

\section{Introduction}

Listening to audio content provides an alternative method for engaging with knowledge that is distinct from visual reading.
Audio listening, often categorized as mobile reading, enables individuals to consume content while performing other tasks, unlike stationary reading, which requires visual focus and one to remain in one place~\cite{TattersallWallin.2020.TimeReadExploringTimespacesa}.
The proliferation of mobile devices and streaming services has facilitated the widespread use of audio content, including audiobooks and podcasts.
\textit{Conversational podcasts} refer to on-demand conversational audio content accessible across various devices, and diverge from traditional definitions~\cite{TattersallWallin.2021.ReadingListeningConceptualisingAudiobook,Bottomley.2015.PodcastingDecadeLifeNew,Farivar.2004.NewFoodIPodsAudio}.
The dialogue format is a distinctive feature of \textit{conversational podcasts} that enhances content understanding~\cite{Muller.2007.ConceptualChangeVicariousLearning,Muller.2008.SayingWrongThingImprovinga,Chi.2017.WhyStudentsLearnMorea,Mayer.2004.PersonalizationEffectMultimediaLearning}.

However, it is impractical for human creators to adapt all niche text content, such as academic papers and government reports, to conversational audio content.
Computational adaptation is crucial to overcome the shortage of niche audio resources.
Text-to-speech (TTS) technology alone can only generate audiobook-like content\footnote{Audiobook-type ``Listen'' features are provided by online libraries, such as Taylor \& Francis or ProQuest~\cite{Taylor&FrancisGroup..TaylorFrancisOnline,ProQuestLLC...ProQuest}.}, which represents a verbatim narration of the written texts~\cite{TattersallWallin.2021.ReadingListeningConceptualisingAudiobook}.
Large language models (LLMs) overcome this limitation by generating conversational scripts from text documents.
Following the release of ChatGPT in late 2022~\cite{OpenAI.2022.IntroducingChatGPT}, systems for automatic document-to-audio conversion have emerged in 2023~\cite{WondercraftInc..[n.d.].FindOutWhoWe,FikaMedia..ThisRecast}.
Saadya and Davis reported a case of medical education using Wondercraft~\cite{Saadya.2021.RevolutionizingPlasticSurgeryEducation,WondercraftInc..[n.d.].AIPoweredAudioStudio}; however, detailed design insights remain sparse.
To date, advanced systems utilizing Large Language Models (LLMs) have improved audio content quality~\cite{Matias.2024.GoogleResearchGoogle2024,Google.2024.Illuminate,Wang.2024.NotebookLMNowLetsYou,Google.2024.NotebookLM,MITLaboratoryforAtomisticandMolecularMechanics.2024.LammmitPDF2Audio,Fitzpatrick.2024.KnowsuchagencyPdfpodcast}.
Given the advancements in LLM, exploring audio-learning methods for niche content is timely.

Therefore, this study investigated the potential of document-to-audio adaptation using LLMs through the design and practical use of a prototype system termed PaperWave\footnote{PaperWave is intended as a supplementary tool alongside reading the original text. Critical reading of the original text remains essential for in-depth understanding.}.
To explore the design considerations of such systems, we focused on research papers in which researchers frequently engaged.
We conducted a field study to understand how researchers may use auditory inputs from research papers.
This approach allowed us to gain insights from the perspective of life-integration.
Given the exploratory nature of this study, researchers from diverse backgrounds were included to capture a wide range of perspectives on the utility of PaperWave across different lifestyles.
Throughout the field study, our overall research process embraced an autobiographical design approach~\cite{Neustaedter.2012.AutobiographicalDesignHCIResearch,Neustaedter.2012.AutobiographicalDesignWhatYou}, which allowed us to document the narrative surrounding the design of both the formats of suitable dialogue and the interfaces for podcast creation and playback.
Finally, we concluded the study with a design workshop to reflect on the process and gather further insights.

The purpose of this study was to provide a nuanced understanding of how a document-to-audio adaptation system using LLMs broadens opportunities for input from niche text documents.
Thus, the following research questions were posed:

\begin{itemize}
    \item RQ1: In what contexts can PaperWave be effectively used?
    \item RQ2: Which aspects of engagement with knowledge can PaperWave enhance, and how?
\end{itemize}

\section{Context of Our Study}

This study was conducted from June 25, 2024, to September 2, 2024, and involved the authors and six collaborators, as detailed in \autoref{tab:participants}.
They have diverse backgrounds, including reading topics, research experience, language proficiency, occupation, and lifestyle.
A4, P2, and P6 have children, and A4 and P6's children include infants.
In this paper, ``we'' refers to the authors (A1--A5, and A6), ``collaborators'' to the study collaborators~(P1--P6), and ``participants'' to both groups.
Citations in the form (P[number], [date]) refer to diaries, and (P[number], WS) refer to workshops.
All the quoted comments were translated from Japanese into English by the authors.

\paragraph{Origin of this Project} On June 25, A3 introduced the concept of PaperWave to A1 and A5, inspired by \cite{Chi.2017.WhyStudentsLearnMorea}. 
The next day, A1 discussed with A2 the potential to treat it as a research project. 
Our initial focus was on enhancing inclusivity and applying media studies through podcast conversions. 
At that time, we were unaware of Google Illuminate\cite{Matias.2024.GoogleResearchGoogle2024, Google.2024.Illuminate}, which meant that our exploration of PaperWave's design proceeded independently of these influences. 
All involved agreed to proceed with the writing of the manuscript, which marked the start of our study.

\paragraph{Sites} Participants primarily used PaperWave in their daily routines in Tokyo and nearby areas. 
They were affiliated with universities or companies located in Tokyo, commuted several times per week, and worked remotely on other days. 
During the August vacation period, participants used PaperWave while returning to their hometowns or on trips. These destinations include South Korea and several Asian countries.

\begin{table*}[t]
  \centering
  \caption{Field study participants and their backgrounds. ``G'' = Gender, ``FL'' = First language, ``Exp'' = Research experience (years), ``WS'' = Workshop Group (``1A'' = day 1, group A), ``Research'' = Proficient for research use, ``Daily'' = Daily conversational level. }
  \label{tab:participants}
  \begin{tabular}{llllllllll}
  \toprule
  \textbf{ID} & \textbf{Age} & \textbf{G} & \textbf{FL} & \textbf{Languages} & \textbf{Occupation} & \textbf{Exp} & \textbf{Reading Topics} & \textbf{Entry} & \textbf{WS} \\
  \midrule
  A1 & 25--29 & M & ja & \begin{tabular}[t]{@{}l@{}}en: Research\end{tabular} & PhD student (3rd) & 6 & \begin{tabular}[t]{@{}l@{}}Maker education\end{tabular} & 6.25 & 2A \\[0.3em]
  A2 & 20--24 & M & ja & \begin{tabular}[t]{@{}l@{}}en: Research\end{tabular} & Master student (2nd) & 5 & \begin{tabular}[t]{@{}l@{}}Design, CSCW\end{tabular} & 6.26 & 1A \\[0.3em]
  A3 & 30--34 & M & ja & \begin{tabular}[t]{@{}l@{}}en: Daily\end{tabular} & \begin{tabular}[t]{@{}l@{}}Business development \\ Product manager\end{tabular} & 5 & \begin{tabular}[t]{@{}l@{}}CS education, \\ Business management\end{tabular} & 6.25 & 1B \\[0.3em]
  A4 & 35--39 & M & ko & \begin{tabular}[t]{@{}l@{}}ja, en: Research\end{tabular} & Lecturer & 15 & \begin{tabular}[t]{@{}l@{}}Digital fabrication\end{tabular} & 8.2 & 2B \\[0.3em]
  A5 & 30--34 & M & ja & \begin{tabular}[t]{@{}l@{}}en: Daily\end{tabular} & Researcher & 9 & \begin{tabular}[t]{@{}l@{}}Lifelong learning, \\ Media studies\end{tabular} & 6.25 & 1B \\
  \midrule
  P1 & 20--24 & M & ja & \begin{tabular}[t]{@{}l@{}}en: Daily\end{tabular} & Master student (1st) & 1 & \begin{tabular}[t]{@{}l@{}}Mental health, \\ Writing with AI\end{tabular} & 8.12 & 2A \\[0.3em]
  P2 & 40--44 & M & ja & \begin{tabular}[t]{@{}l@{}}en: Daily\end{tabular} & Researcher & 1 & \begin{tabular}[t]{@{}l@{}}CS education, \\ Machine learning\end{tabular} & 8.13 & 2A \\[0.3em]
  P3 & 20--24 & M & ja & \begin{tabular}[t]{@{}l@{}}en: Daily\end{tabular} & Master student (1st) & 1 & \begin{tabular}[t]{@{}l@{}}Game commentary, \\ Input methods\end{tabular} & 8.12 & 1A \\[0.3em]
  P4 & 20--24 & F & ja & \begin{tabular}[t]{@{}l@{}}en: Research, \\ de: Daily\end{tabular} & Master student (1st) & 1 & \begin{tabular}[t]{@{}l@{}}Lifelogs, Slow-tech, \\ Behavior change\end{tabular} & 8.15 & --- \\[0.3em]
  P5 & 20--24 & F & ja & \begin{tabular}[t]{@{}l@{}}en: Daily\end{tabular} & Undergraduate (4th) & >1 & \begin{tabular}[t]{@{}l@{}}Psychology, \\ Behavior change\end{tabular} & 8.15 & 2B \\[0.3em]
  P6 & 40--44 & F & ko & \begin{tabular}[t]{@{}l@{}}ja: Research, \\ en: Daily\end{tabular} & Professor & <10 & \begin{tabular}[t]{@{}l@{}}Design\end{tabular} & 8.19 & 2B \\
  \bottomrule
  \end{tabular}
\end{table*}


\paragraph{Who We Are} We have disclosed our background to address the potential biases in our findings. 
Our primary interest lies in learning and design, and not in the technology of LLMs, which directs our focus to user experiences. 
Our optimism regarding technology might have led to the underestimation of the challenges of introducing PaperWave. 
A1 took a year off university partly owing to stress from literature reviews. 
Because of this experience, A1 is seeking alternative methods for reading papers, which may have biased him toward justifying PaperWave. 
A3, a product manager, reads academic papers for work and hobbies, offering diverse perspectives that differ from typical researchers needs. 
To counterbalance our enthusiasm for PaperWave, we included A5, a potential non-user who hesitated to use PaperWave because of its focus on the paper's body~(\ref{subsec:result-body}) and because he was satisfied with the current text-based methods for reading papers.
Professor A6, who has worked in pioneering interdisciplinary research areas, contributed to the study as a supervisor by offering comments, primarily on the workshop.
Additionally, the participants had a first language other than English, which may have significantly influenced the results, allowing them to engage with English papers in their first language.

\section{Method}

\subsection{Field Study}

A field study was conducted to explore the potential of PaperWave in real-world settings. 
All collaborators are from the two laboratories that the authors belong to and were recruited them through lab Slack workspaces.
The collaborators began with a 30-minute entry interview to discuss their reading habits and challenges. 
The participants were free to use PaperWave for research or personal interest. 
They documented their experiences in diaries on Slack. 
The authors shared a single channel, whereas the collaborators had private channels to maintain confidentiality.
Ethical considerations, such as providing informed consent and ensuring privacy, were followed.
Local regulations do not require a formal ethics reviews in field studies involving laboratory members only.
In total, 1259 messages were recorded, including automatically generated messages such as new member alerts.

\subsection{Autobiographical Design}

This research, including the field study, is grounded in an autobiographical design~\cite{Neustaedter.2012.AutobiographicalDesignWhatYou,Neustaedter.2012.AutobiographicalDesignHCIResearch}, a first-person study approach that shows an increasing sense of maturity in the HCI community~\cite{Desjardins.2021.IntroductionSpecialIssueFirstPerson}.
This method provides a detailed and nuanced understanding of the design space by incorporating real-world experiences in which the system developers themselves are users~\cite{Neustaedter.2012.AutobiographicalDesignWhatYou}.
The authors face challenges in reading papers, such as language barriers, balancing research with household duties and childcare, and difficulty engaging with unfamiliar fields for interdisciplinary research.
Amidst these challenges, the authors explored the design of PaperWave through \textit{genuine usage}~\cite{Neustaedter.2012.AutobiographicalDesignHCIResearch} along with collaborators.
As demonstrated in \cite{Mellis.2014.DoItYourselfCellphonesInvestigationPossibilities, Desjardins.2016.LivingPrototypeReconfiguredSpace}, this approach allows for the examination of designs closely tied to everyday life through intensive use~\cite{Neustaedter.2012.AutobiographicalDesignWhatYou}.
Moreover, the PaperWave case, which utilizes the latest LLMs, exemplifies \textit{early innovation} facilitated by \textit{fast tinkering} with this approach~\cite{Neustaedter.2012.AutobiographicalDesignHCIResearch}.
Although the two-month usage period in this study was brief, we transcended a simple field study by actively incorporating a first-person approach to document the design process and researchers' influence, which might otherwise remain hidden~\cite{Neustaedter.2012.AutobiographicalDesignHCIResearch,Desjardins.2021.IntroductionSpecialIssueFirstPerson,Ellis.2011.AutoethnographyOverview}.
Following prior research advice, we used PaperWave with a non-users (A5, A6) and secondary users (P1--P6)~\cite{Neustaedter.2012.AutobiographicalDesignHCIResearch}.
A1 and A2 developed PaperWave, and the authors held weekly meetings to discuss improvements and issues.
The source code was hosted on GitHub, and improvements were managed through issues features~\cite{GitHub.[n.d.].GitHub}.

\subsection{Design Workshop}

The design workshop aimed to deepen the participants' understanding of their experiences with PaperWave and explore new possibilities through creative activities. 
Held online on September 1st and 2nd, sessions were facilitated by A1 and A3, respectively. 
The contents of the two workshops were the same and the participants attended either one or the other.
Participants used Zoom and Figma FigJam for collaborative brainstorming, communicating in Japanese. 
The workshop comprised three activities. 
\textit{Activity 1} focused on sharing personal uses for PaperWave. 
\textit{Activity 2} explores the types of knowledge engagement facilitated by PaperWave. 
\textit{Activity 3} involved discussing improvements and the future design of PaperWave. 
The ideas generated by participants were based on diary entries from the field study.
P4 could not attend the workshop due to time constraints and only the diary entries were used for analysis.
The workshop format, rather than group interviews, was chosen to foster creative exploration and deeper understanding through discussions. 

\subsection{Analysis of the Workshop and Field Study}

The workshop and field study data were analyzed using MAXQDA. 
For RQ1, A1 focused on coding diary entries from the field study, as it was challenging to capture all the experiences during the workshop. 
A1 coded diary text to identify situations in which PaperWave was used, integrating similar codes until they converged. 
For RQ2, the analysis involved bottom-up coding of sticky notes from \textit{Activity 2} while maintaining the hierarchical structure of headings and ideas. 
A1 coded the headings in vivo and merged similar codes. 
Participants continued to use PaperWave beyond the study period and wrote diaries, which potentially influenced the authors' analysis.
The diary entries after September 2nd were not included in the formal analysis.

\section{PaperWave}

\subsection{Design Rationale}

Drawing on the insights gained through the autobiographical design, we established the initial design rationale for PaperWave. 
This rationale outlines the potential benefits and guiding principles of the design.

\paragraph{Design Rationale for Conversation Content and Format} \ref{format:interview-style} allows users to catch up on content even if they miss parts. \ref{format:outline} enhances comprehension and engagement. \ref{format:accuracy} includes the exclusion of unrelated content, such as commercials or announcements.

\aptLtoX[graphic=no,type=html]{\begin{enumerate}
    \item[FM1] \label{format:interview-style} \textbf{Adopt an interview-style speech}, including questions and rephrasing by the program host. 
    \item[FM2] \label{format:outline} \textbf{Present an outline} of the topics covered in the podcast at the beginning of the program.
    \item[FM3] \label{format:accuracy} \textbf{Convey the content written in the paper accurately} without omissions or additions.
\end{enumerate}}{\begin{enumerate}[label={\lbrack FM\theenumi \rbrack}, ref={\lbrack FM\theenumi \rbrack}]
    \item \label{format:interview-style} \textbf{Adopt an interview-style speech}, including questions and rephrasing by the program host. 
    \item \label{format:outline} \textbf{Present an outline} of the topics covered in the podcast at the beginning of the program.
    \item \label{format:accuracy} \textbf{Convey the content written in the paper accurately} without omissions or additions.
\end{enumerate}}

\paragraph{Design Rationale for Podcast Creation and Playback Interface}
\ref{if:responsive} enables \textit{mobile reading}. \ref{if:duration} allowed users to adapt podcasts to their time constraints. \ref{if:bgm} provided a fresh and engaging atmosphere. \ref{if:sharing} encourages collaboration and exposure to various topics.

\aptLtoX[graphic=no,type=html]{\begin{enumerate}
    \item[ IF1] \label{if:responsive} \textbf{Offer a responsive player} interface compatible with both desktop and mobile devices.
    \item[ IF2] \label{if:duration} \textbf{Allow users to specify the duration} of an adapted podcast.
    \item[ IF3] \label{if:bgm} \textbf{Play BGM} along with the speech.
    \item[ IF4] \label{if:sharing} \textbf{Enable sharing} of created episodes with colleagues.
\end{enumerate}}{\begin{enumerate}[label={\lbrack IF\theenumi \rbrack}, ref={\lbrack IF\theenumi \rbrack}]
    \item \label{if:responsive} \textbf{Offer a responsive player} interface compatible with both desktop and mobile devices.
    \item \label{if:duration} \textbf{Allow users to specify the duration} of an adapted podcast.
    \item \label{if:bgm} \textbf{Play BGM} along with the speech.
    \item \label{if:sharing} \textbf{Enable sharing} of created episodes with colleagues.
\end{enumerate}}

\subsection{System Overview}

PaperWave has three features: \textit{Recording}, \textit{Episodes}, and \textit{Channels}.
On the \textit{Recording} page, users upload a PDF and customize the podcast by specifying the episode title, duration, language, and the LLM model for script generation.
This customization allows users to tailor the podcast episodes according to their needs.
The generated podcasts are accessible through the responsive \textit{Episodes} page~(\autoref{fig:teaser}~(A)), facilitating listening in various contexts such as commuting or household tasks.
In addition, the \textit{Channels} page allows users to explore and listen to episodes created by colleagues, fostering a collaborative environment and exposure to a wider range of topics.
Because the implementation of the PaperWave system is not the main contribution of this case study, further details on the features and implementations are provided in \autoref{sec:appendix-system-details}.

\subsection{Design Process of Three PaperWave Prototype}

The PaperWave design process comprised three major phases: manual generation using ChatGPT, automatic generation using a command line interface (CLI), and a web application.
The findings and design implications of these three phases are as follows:

\subsubsection{First Phase: Manual Generation using ChatGPT}

In the first phase, conducted from June 25 to July 9, we explored prompts that generated podcast scripts from PDFs by manually inputting prompts into ChatGPT.
The script was converted into speech using a Python script.
We realized the importance of making LLMs with different roles cooperate, as described in \autoref{subsec:implementation}.
This approach was motivated by the need to create long programs (\textit{e.g.} 15 min) for activities such as aerobic exercise (A1, June 26), as single LLMs could not generate content of sufficient length.
We also experimented with a news format (A3, June 28) but focused on the interview format because of its effectiveness in the dialogue format~\ref{format:interview-style}.

\subsubsection{Second Phase: Command Line Interface (CLI) for Automatic Generation}

The development of a commandline interface for podcast generation began on July 1 and continued until August 21, when the web application became operational.
Iterative adaptation and listening revealed several effective and problematic features.
The key findings include the benefits of presenting a podcast outline at the beginning~\ref{format:outline}, which enhances user engagement and comprehension (A3, August 4 and A2, August 9).
Avoiding unrelated content~\ref{format:accuracy} is essential for maintaining a focus on the core material (A1, August 3).
The influence of background music (BGM)~\ref{if:bgm} has been noted because it provides a fresh and engaging listening experience.
Without BGM, the audio was perceived as too quiet (A1, August 4), and different BGM tracks offered a refreshing change (A1, August 13).
This also aids in recognizing the transitions between successive podcast episodes (P2, August 13).
Tailoring the podcast duration to match the paper length~\ref{if:duration} is beneficial while needing to matching the duration with the paper length~\ref{if:duration} (A1, July 28).
Enabling sharing with colleagues~\ref{if:sharing} is advantageous, as it allows users to conveniently engage with papers shared by colleagues, particularly when they are not in a position to focus on reading (A1, August 18).
Finally, accessibility using smartphones is crucial for \textit{mobile reading}~\ref{if:responsive} (\textit{e.g.}, A2, August 2).

\subsubsection{Third Phase: Web Application}\label{subsec:design-third-phase}

Based on the findings of the first and second phases, we developed a web application.
The findings from this phase are described in the next section as result of a field study.

\section{Findings from Field Study and Design Workshop}

\subsection{RQ1: PaperWave Enables Mobile Reading of Papers}\label{subsec:result-mobile-reading}

The participants used PaperWave in diverse contexts and recorded their experiences in diaries. 
\textit{While traveling}, nearly all the participants listened to the podcast episodes. 
They used public transportation, such as trains and buses~(A1, A2, A3, A4, P1, P2), walked~(A1 (\autoref{fig:teaser}~(C)), P2, P5), and drove~(A1). 
\textit{While performing activities}, participants engaged in podcast episodes during housework~(A1 (\autoref{fig:teaser}~(B)), P2, P5, P6), childcare~(A4, P6), preparing for going out~(A1, P3), paperwork~(A1), and video editing~(P3). 
\textit{Break time} provided another opportunity to listen. 
Unlike activity times, the break time focused solely on listening. 
Participants listened while waiting~(A3, A5), before sleeping~(A1, A4), eating~(A1), sitting in a massage chair~(P6), and during work breaks~(P5).

PaperWave facilitates engagement with research papers in contexts where visual reading was impractical. 
However, their usefulness is limited to certain scenarios. 
Participants reported challenges in processing research papers through podcast listening. 
For instance, one participant noted, ``If I do other tasks while listening, I won't understand it'' (P1, WS). 
There were instances in which the participants struggled to fully engage with the content. 
One participant mentioned,  ``I started doing other work while listening during breaks, and I didn't listen from the middle'' (P5, August 22). 
Another participant observed, ``If I listen during sleepy times, there is also a pattern of `falling asleep' '' (A1, August 27).

\subsection{RQ2: PaperWave Changes the Participants Engagement with Knowledge}

\subsubsection{Lowers the Barrier to Engaging with Papers Compared to Text Documents}\label{subsubsec:lowers-barrier}

Creating podcasts in the participants' first language lowered the barrier to engaging with research papers, especially for those less proficient in English. 
As P1 expressed, ``the hurdle to deciding whether or not to engage with is lowered. I feel like I can try listening to it just for now.'' (P1, WS). 
The transition to audio-enabled \textit{mobile reading} allowed the participants to engage without needing to stay on a computer (unknown participant from group 2B, WS). 
This ``reduced the mental burden of dedicating specific time blocks'' for reading (P6, WS). 
Additionally, this shift allowed for immediate access to papers shared by others (A1, WS) and encouraged the exploration of a broader range of topics than usual (A1, A2, A4, P3, WS). 
P3 noted, ``I can casually listen to that seems interesting when expanding my field'' (P3, WS). 
One participant also found that enjoying papers as entertainment facilitated casual engagement, making academic content shared by others more approachable and enjoyable (unknown participant from group 2B, WS).

\subsubsection{Different Emphasis on Information Compared to Reading Text Document}\label{subsubsec:different-emphasis}

When participants listened to papers as podcasts, their attention shifted to different aspects of the content compared with reading text documents.
For instance, A2 remarked that ``information often overlooked in text became significant when listened'' (WS).
A3 explained this shift by stating, ``I can read papers in a more balanced way'' (A3, August 4).
P2 found the conversational format beneficial for understanding, noting it ``helped me to understand it better'' (P2, WS).
This auditory format also facilitats multitasking, allowing P2 to write a budget application while searching for citation information.
He would ``playback the paper that I think had the information I wanted to cite, and when I found it, I would go to the paper document,'' thus enabling simultaneous writing and information retrieval (P2, WS).
Conversely, P6 reported that listening to the podcast without a clear purpose sparked new research ideas such as using certain messages and frameworks in her work (P6, WS).
Furthermore, the podcast's format allowed participants like A1 and A2 to feel a personal connection, as if ``I've become a fan of the author. Even if I met the author at a conference, I don't think I would feel like I was meeting [AUTHOR] for the first time” (A2, WS, the author's name is omitted).


\subsubsection{Focused Mainly on the Paper's Body}\label{subsec:result-body}

In the PaperWave podcast episodes, references were seldom mentioned, with most discussions centered on the paper's body, such as methods and results.
This approach allowed the participants to gain a rough understanding of a single paper.
P3 noted that ``using PaperWave to input many articles in summary'' (WS) was effective.
A2 observed that the research flow was ``easier to follow than the (text) abstract'' (WS, words in parentheses are additional notes by the authors).
Opinions varied depending on whether the focus was beneficial or limited.
A3 and A5 illustrate this difference.
A5 explained, ``I feel the difference comes from whether a person wants to read the research results (A3), or the introduction (A5)'' (WS, speech noted by A1).
A3, a business practitioner, reads papers to explore daily interests and advise colleagues (A3, WS).
Conversely, A5, a university researcher with a sociological background, valued understanding the research context, emphasizing its importance over individual findings.
Thus, A5 was reluctant to use PaperWave (WS, speech noted by A1).
Other participants also expressed concerns about the lack of contextual information (A1, A2, P3, WS).

\subsubsection{Lacking Information and Accuracies}

During the workshop, the participants identified two main issues: the absence of visual information and concerns about inaccuracies. 
The lack of visual information is an inherent limitation of audio content, as noted for A1, A2 and P1 (WS). 
This absence makes it challenging to grasp the content fully, particularly in fields such as design, where visual elements are crucial. 
Concerns regarding inaccuracies were particularly pronounced when engaging with content outside one's area of expertise. 
A2 expressed a fear of encountering inaccurate information in such contexts (WS). 
In addition, A4 who listened to a paper authored by himself, remarked, ``When the content is \textit{almost at the edge of hallucination} (ギリギリハルシネーションに近い), it raises the question of whether I wrote it badly or if LLM's analytical abilities are deficient'' (A4, August 3). 
A1 agreed, explaining that sometimes the emphasis on the content shifts slightly, leading to a nuance that differs from the author's intention (A1, August 3). 
These issues highlight the challenges of ensuring the accuracy and completeness of audio-based content delivery.


\section{Discussions and Lessons Learned}

As demonstrated in \ref{subsec:result-mobile-reading}, PaperWave facilitates \textit{mobile reading}, enabling participants to engage with research papers across various contexts, albeit with certain challenges.
This transition to audio content parallels the case to audiobooks~\cite{TattersallWallin.2020.TimeReadExploringTimespacesa,TattersallWallin.2022.AudiobookRoutinesIdentifyingEveryday}.
While Tattersall-Wallin and Nolin examined audiobook engagement as a leisure activity, suggesting that reading occurs during working hours~\cite{TattersallWallin.2020.TimeReadExploringTimespacesa}, our study indicated the opposite; engagement with papers related to professional activities may be integrated into private life.
Moreover, the participants' activities often interacted with podcast listening.
For example, P2 managed to write a budget application while listening to podcasts, showcasing a unique form of multitasking.
Another example is A3, who, while listening to ``ARECA: A Design Speculation on Everyday Products Having Minds''~\cite{Cho.2023.ARECADesignSpeculationEveryday}, imagined the implications of vending machines having minds during a walk (July 29th).
He applies this concept to the design of urban vending machines.
This example underscores the benefits of \textit{mobile reading}, which allows engagement with academic papers while existing in the physical world.
These findings highlight, unlike reading support systems for text documents that focus solely on content (\textit{e.g.} \cite{August.2023.PaperPlainMakingMedical,Fok.2023.ScimIntelligentSkimmingSupport,Chang.2023.CiteSeeAugmentingCitationsScientific,Kim.2023.PapeosAugmentingResearchPapers,Lee.2024.PaperWeaverEnrichingTopicalPaper}), the necessity for document-to-audio systems to be designed considering of the interaction context with the surrounding world.

The adaptation of text to audio has demonstrated its potential to lower barriers to engaging with academic papers (\ref{subsubsec:lowers-barrier}) and alter comprehension (\ref{subsubsec:different-emphasis}).
This aligns with previous research indicating that conversational content aids understanding~\cite{Muller.2007.ConceptualChangeVicariousLearning,Muller.2008.SayingWrongThingImprovinga,Chi.2017.WhyStudentsLearnMorea,Mayer.2004.PersonalizationEffectMultimediaLearning}.
The availability of an overview makes it easier to access papers in unfamiliar fields, which is consistent with research on skimming support systems~\cite{Fok.2023.ScimIntelligentSkimmingSupport}; however, in the case of a document-to-audio system, it does not necessarily mean that the content can be accessed in a short time.
However, several interesting cases have been reported.
The reduction in barriers is not solely due to language changes but also because \textit{mobile reading} diversifies listening contexts.
The passive reception of information as audio progresses can shift the emphasis.
Moreover, the speech format of the AI author may influence the user's sense of intimacy with the author.
In addition, a participant expressed concern about the choice of speakers, stating, ``I think there would be a sense of resistance to the audio that was created as if I was the one speaking, like `narration rights' such as `portrait rights{'} (肖像権ならぬ、語り方権のような自分が語ってるとして作られた音声への抵抗感って出てきそう)'' (P2, July 29) that illustrates the necessity to discuss on the choice of the speakers in the audio content.
While academic reading typically focuses on content comprehension, shifts in focus and emotional impacts should be considered when designing document-to-audio systems.
A a vast and unexplored design space remains to create conversational podcast formats.

In this study, we employed a first-person approach that facilitated reflection on our engagement with papers. 
As noted in \ref{subsec:result-body}, focusing on the body of a paper is a pivotal factor in assessing the usefulness of document-to-audio adaptation systems. 
This focus appears to be a common trait among such systems~\cite{Google.2024.NotebookLM,Google.2024.Illuminate,FikaMedia..ThisRecast}, as observed by the authors. 
Here, A3 and A5 delve deeper into autobiographical reflections.
A3 believed that if knowledge is generated by a researcher, even partial extraction and application in daily life is valuable.
He views papers not only as complete works but also as components, prioritizing inspiration over potential inaccuracies due to adaptation. 
Conversely, A5 appreciates the IMRaD structure of research papers and prefers to engage with them in their original text format, even if the quality of PaperWave improves. 
This dichotomy underscores the varied perspectives of document-to-audio systems, reflecting the broader implications of transitioning to audio content as explored in the PaperWave design process.
By exploring new forms of engagement with research papers through an autobiographical design, the personal attitudes of each researcher towards how they wanted to engage with knowledge were highlighted. 
Gaining such deep insights through a first-person approach is the significance of this research.

\section{Conclusion}

This study demonstrated the transformative potential of LLMs in reshaping how we engage with research papers.
By converting text documents into conversational podcasts, PaperWave offers a more accessible and flexible means for interacting with knowledge. 
These findings underscore the diverse contexts in which PaperWave can be effectively utilized. 
The design process highlights the necessity of considering listener interactions with their surrounding environment when developing document-to-audio systems.

However, this study has certain limitations. 
The findings and lessons cannot be generalizable. 
Additionally, the diary method, which involves sending Slack messages, may not have fully captured instances in which PaperWave was not used. 
Although non-use reports were intentionally included, a more routine diary-keeping method is required for a detailed examination.
Furthermore, the lack of personalization, with all participants engaging in podcasts generated from the same prompts, presents a limitation and opportunity for further exploration. 
Despite these limitations, we successfully captured the nuanced relationships between text documents, audio content, and user engagement with knowledge.

In \textit{Activity 3} of the workshop, proposals were made to address the challenges faced by the participants. 
These include the potential to integrate reading support technologies for text papers, such as content recommendations and multimodal augumentation of text documents~\cite{Lee.2024.PaperWeaverEnrichingTopicalPaper,Kim.2023.PapeosAugmentingResearchPapers}, into PaperWave. 
The application of insights from the HCI community to document-to-audio systems is a promising future direction. 
We hope that this study provides unique insights into document-to-audio adaptation and engagement with the rapidly evolving LLM technology, stimulating further discussions within the HCI community.

\begin{acks}
This work was supported by Nakayama Future Factory and JSPS KAKENHI Grant Number JP22KJ1010. 
We would like to thank the collaborators of the field study and the design workshop for their valuable insights and feedback.
We also thank the anonymous reviewers for their constructive comments.
The original draft of the figure descriptions for accessibility was written by ChatGPT-4o and then revised by the authors.
\end{acks}

\bibliographystyle{ACM-Reference-Format}
\bibliography{references}

\appendix

\section{PaperWave System Details}\label{sec:appendix-system-details}

A system walk-through and details of implementation are provided to help readers learn more about the PaperWave system and understand the case description in more depth.

\subsection{System Walk-Through: an Example User Scenario}

By presenting a scenario in which a user engages with PaperWave, we provide an overview of its features. The example given in this section is a scenario in which a user wants to listen to paper podcasts on PaperWave while commuting to a university by foot and train. Before preparing, the user uploads the paper to PaperWave and listens to the paper’s podcast on their smartphone while walking. After boarding the train, the user selects an episode made by another user to listen to. This series of scenarios showcases the \textit{recording} features, \textit{episode playback} features, and channels list features of PaperWave.

\subsubsection{\textit{Recording}}

The user uploads the PDF file and generates a conversational podcast. Technically, a user can upload and adapt multiple research papers into a single podcast episode; however, no user has adapted multiple papers simultaneously. In the PaperWave web application, this feature is termed \textit{recording}, and is used as a metaphor for podcast production.

\begin{figure*}[ht]
  \centering
  \includegraphics[width=0.7\linewidth]{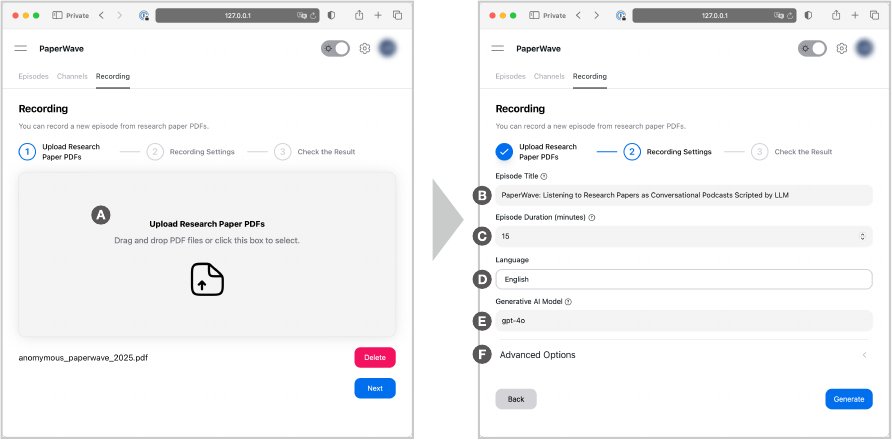}
    \caption{\textit{Recording} page.
    Users can adapt research paper PDFs into conversational podcasts using the interface shown in this figure (this adaptation is referred to as \textit{recording} in the app).
    (A) Upload a PDF file to adapt.
    (B) Enter the title of the episode to \textit{record}.
    (C) Set the duration of the episode.
    (D) Choose the language of the speakers in the podcast episode.
    (E) Choose the LLM model to generate the script.
    (F) Advanced options include episode description, keywords, and cover image URL.}
  \label{fig:4-2-recording-page}
  \Description{PaperWave Recording Interface
The figure shows two screenshots of the PaperWave app interface used for creating podcast episodes from research papers.
- Left Image: The initial "Recording" screen of the PaperWave app is displayed. The screen prompts the user to "Upload Research Paper PDFs" by dragging and dropping PDF files or clicking the box to select files. The file "anonymous_paperwave_2025.pdf" has been uploaded, and two buttons, "Delete" (in red) and "Next" (in blue), are visible at the bottom.
- Right Image: The next step, "Recording Settings," is shown. The user has filled out the "Episode Title" as "PaperWave: Listening to Research Papers as Conversational Podcasts Scripted by LLM." Other settings include "Episode Duration (minutes)" set to 15, "Language" set to English, and the "Generative AI Model" set to "gpt-4o." An "Advanced Options" section is also visible, and a "Generate" button is displayed at the bottom right.}
\end{figure*}

First, the user uploads a PDF file to record (\autoref{fig:4-2-recording-page}~(A)). 
Second, the user enters the title of the episode (\autoref{fig:4-2-recording-page}~(B)). 
Third, the user sets the duration of the episode (\autoref{fig:4-2-recording-page}~(C)).
The user specifies 15 minutes because their home and the nearest station are about a 15-minute walk away.
Users can specify the duration of the episode to suit their listening situation. 
Fourth, the user chooses the language of the speakers in the podcast episode (\autoref{fig:4-2-recording-page} (D)).
The user can select a language that they prefer, regardless of the language in which the source research paper is written.
Current implementation allows users to select English, Japanese, and Korean, which were the preffered languages for the authors and participants.
Additionally, the user can choose the LLM model from the list of OpenAI's models to generate the script (\autoref{fig:4-2-recording-page} (E)).
There are more advanced options, such as episode description, keywords, and cover image URL, as podcast metadata (\autoref{fig:4-2-recording-page} (F)).
Finally, the user clicks the ``Generate'' button to start the \textit{recording} process.

\subsubsection{Episodes Playback}

\begin{figure*}[t]
  \centering
  \includegraphics[width=0.7\linewidth]{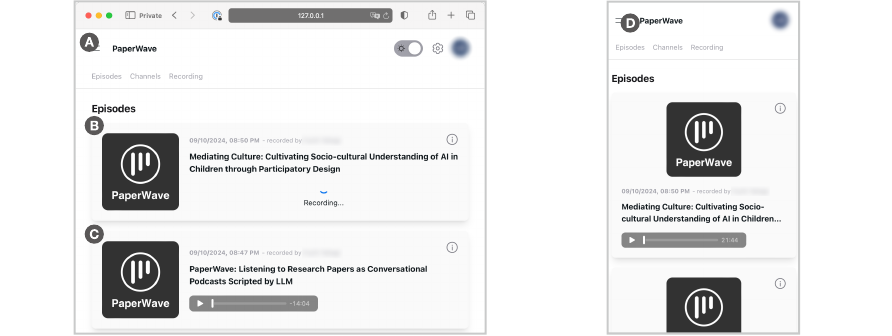}
    \caption{Episodes page shows a list of recorded podcasts.
    (A) Episodes page on large screens.
    (B) After the user has input the PDF, the episode will be displayed with a recording status until the generation is complete.
    Depending on the duration specified by the user, the recording will take about five minutes.
    (C) When the recording is complete, the player interface will be displayed.
    Users will use this interface to playback episodes.
    (D) Episodes page on mobile devices.
    All pages of PaperWave support responsive design and can be accessed from mobile devices.
    Users can listen to episodes in various locations while doing everyday tasks, such as chores or traveling, with their mobile devices.}
  \label{fig:4-2-episodes-page}
  \Description{PaperWave Episodes Page and Mobile View
This figure displays two screenshots of the PaperWave app interface showing the "Episodes" page.
- Left Image: The desktop view of the "Episodes" page in the PaperWave app. The page lists multiple podcast episodes, including one titled "Mediating Culture: Cultivating Socio-cultural Understanding of AI in Children through Participatory Design," which is currently in the "Recording..." state. Another episode, "PaperWave: Listening to Research Papers as Conversational Podcasts Scripted by LLM," is displayed with a playback bar indicating a duration of 14 minutes and 4 seconds.
- Right Image: The mobile view of the "Episodes" page, displaying the two episodes. The episode "Mediating Culture: Cultivating Socio-cultural Understanding of AI in Children through Participatory Design" shows a playback duration of 21 minutes and 44 seconds. The title and playback controls for the second episode is not shown as the contents are placed outside of the visible area.}
\end{figure*}

After the user has inputted the PDF and started the \textit{recording} job, the episode will be displayed as a recording status (\autoref{fig:4-2-episodes-page} (B)).
Depending on the duration specified by the user, the recording will take about five minutes.
During this processing time, the user can get ready.
When the recording is complete, the player interface will appear (\autoref{fig:4-2-episodes-page} (C)).
Since the user is going for a walk to the station, they will access PaperWave from their mobile device to listen to the \textit{recorded} episode.
PaperWave supports both desktop and mobile devices (\autoref{fig:4-2-episodes-page} (A) and (D)).
Not only the episodes page, but all pages are responsive. Thanks to this multi-device support, users can listen to episodes in various locations while doing everyday tasks, such as chores or traveling, with their mobile devices.

\subsubsection{\textit{Channels} List to Explore Colleagues' Episodes}

\begin{figure*}[t]
  \centering
  \includegraphics[width=0.7\linewidth]{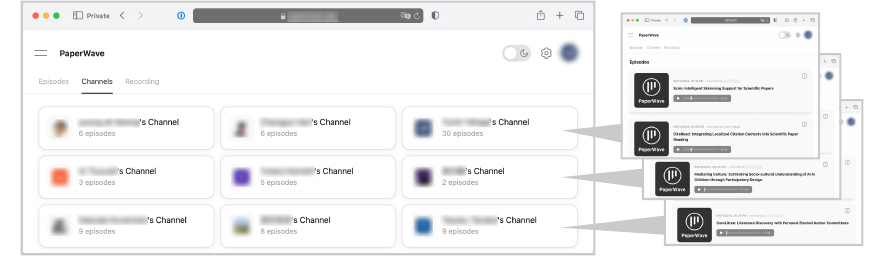}
  \caption{
  \textit{Channels} page shows a list of episodes created by colleagues.
  Users can select a \textit{channel} and visit the colleague's episodes page to listen to the episodes \textit{recorded} by the colleague.
  }
  \label{fig:4-2-channels-page}
  \Description{PaperWave Channels Interface
The figure illustrates the "Channels" page of the PaperWave app and a detailed view of available episodes within selected channels.
- Left Image: The "Channels" page in the PaperWave app shows a list of different channels created by nine users. Each channel displays the user's name, profile picture, and the number of episodes available in that channel. User names and profile pictures are blurred.
- Right Image: An expanded view of selected channels, displaying a list of podcast episodes within each channel. Episodes are listed with titles, timestamps, and playback controls. Examples include "Scim: Intelligent Skimming Support for Scientific Papers" with a duration of 5 minutes and 35 seconds, and "CiteRead: Integrating Localized Citation Contexts into Scientific Paper Reading" with a duration of 6 minutes and 34 seconds.}
\end{figure*}

By the time the user arrives at the station, they will have finished listening to the episode the user recorded themself.
Then, the user will access the \textit{channels} page to explore episodes created by colleagues (\autoref{fig:4-2-channels-page}). 
This function allows users to listen to episodes recorded by colleagues who share their interests, and it facilitates listening to podcasts on a wide range of topics.
A user found an episode of a paper that a colleague had shared on Slack yesterday and listened to it on the train.

\subsection{Implementation}\label{subsec:implementation}

PaperWave system is consisits of a backend PDF-to-podcast adaptation script and a web application.

\subsubsection{Backend Script (\textit{PaperWave CLI})}

The backend script (\textit{PaperWave CLI}) is responsible for converting research papers into conversational podcasts.
The script can be run from the command line interface (CLI) or the web application.
The backend script is implemented in TypeScript and uses OpenAI's APIs for the language model (LLM), PDF processing, and text-to-speech (TTS) technology.

\begin{figure*}[t]
  \centering
  \includegraphics[width=\linewidth]{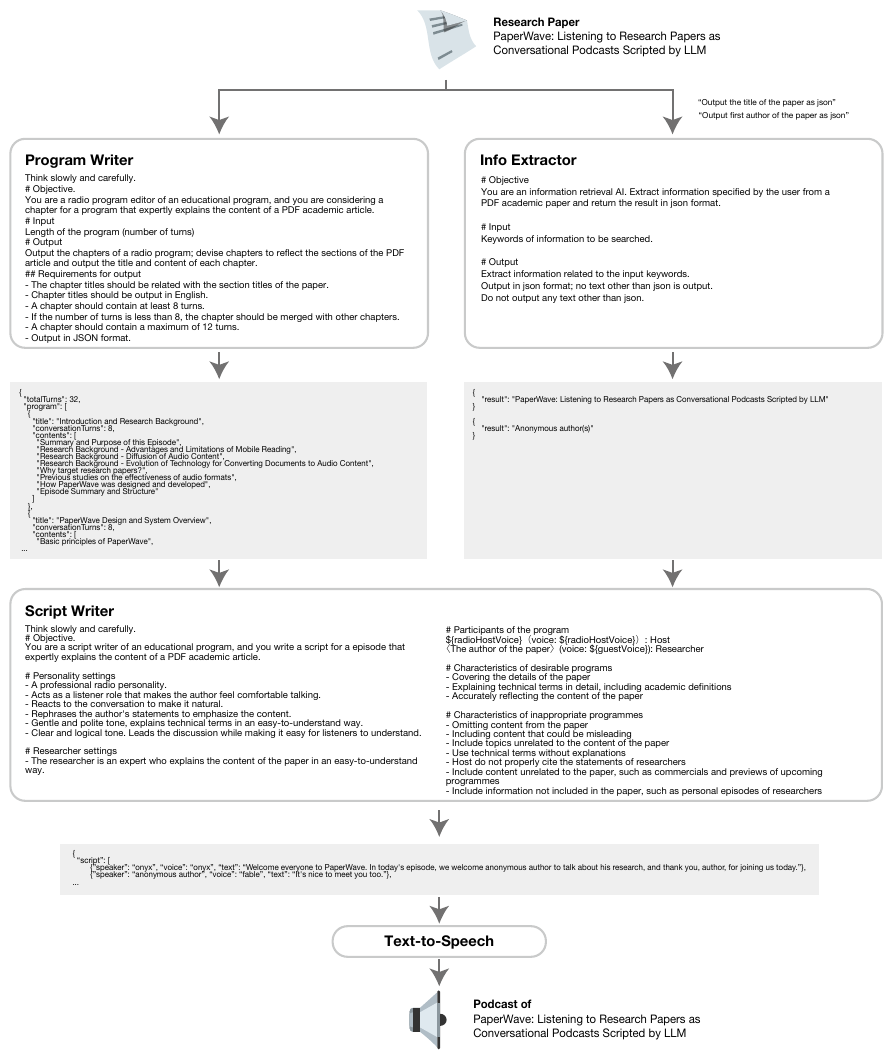}
    \caption{PaperWave CLI implementation. The boxed texts show the instructions for the LLM for adaptation. Process from the input of the PDF to the audio of the podcast is illustrated.}
    \Description{Workflow for Converting a Research Paper into a Podcast Using PaperWave
This figure illustrates the step-by-step process used by PaperWave to convert a research paper into a podcast format, using a combination of program writers, information extraction, script writing, and text-to-speech technology.
1. Research Paper: The process begins with the input of a research paper titled “PaperWave: Listening to Research Papers as Conversational Podcasts Scripted by LLM.”
2. Info Extractor:
\# Objective
You are an information retrieval AI. Extract information specified by the user from a PDF academic paper and return the result in json format. 
\# Input
Keywords of information to be searched. 
\# Output
Extract information related to the input keywords.
Output in json format; no text other than json is output.
Do not output any text other than json.Example outputs include the title of the paper and the names of the authors.
3. Program Writer:
Think slowly and carefully.
\# Objective.
You are a radio program editor of an educational program, and you are considering a chapter for a program that expertly explains the content of a PDF academic article.
\# Input
Length of the program (number of turns)
\# Output
Output the chapters of a radio program; devise chapters to reflect the sections of the PDF article and output the title and content of each chapter.
\#\# Requirements for output
- The chapter titles should be related with the section titles of the paper.
- Chapter titles should be output in English.
- A chapter should contain at least 8 turns.
- If the number of turns is less than 8, the chapter should be merged with other chapters.
- A chapter should contain a maximum of 12 turns.
- Output in JSON format.
4. Script Writer:
Think slowly and carefully.
\# Objective.
You are a script writer of an educational program, and you write a script for a episode that expertly explains the content of a PDF academic article.
\# Personality settings
- A professional radio personality.
- Acts as a listener role that makes the author feel comfortable talking.
- Reacts to the conversation to make it natural.
- Rephrases the author's statements to emphasize the content.
- Gentle and polite tone, explains technical terms in an easy-to-understand way.
- Clear and logical tone. Leads the discussion while making it easy for listeners to understand.
\# Researcher settings
- The researcher is an expert who explains the content of the paper in an easy-to-understand way.
\# Participants of the program
radioHostVoice（voice: radioHostVoice）: Host
〈The author of the paper〉(voice: guestVoice): Researcher
\# Characteristics of desirable programs
- Covering the details of the paper
- Explaining technical terms in detail, including academic definitions
- Accurately reflecting the content of the paper
\# Characteristics of inappropriate programmes
- Omitting content from the paper
- Including content that could be misleading
- Include topics unrelated to the content of the paper
- Use technical terms without explanations
- Host do not properly cite the statements of researchers
- Include content unrelated to the paper, such as commercials and previews of upcoming programmes
- Include information not included in the paper, such as personal episodes of researchers
5. Text-to-Speech:
6. Podcast Output:
The final output is a podcast episode titled “PaperWave: Listening to Research Papers as Conversational Podcasts Scripted by LLM,” created using the text-to-speech technology based on the structured script.
    }
  \label{fig:4-3-implementation}
\end{figure*}

PaperWave CLI, as depicted in \autoref{fig:4-3-implementation}, involves three primary assistants: the \textit{Program Writer}, the \textit{Info Extractor}, and the \textit{Script Writer}.
The \textit{Program Writer} is responsible for crafting the episode's structure by generating chapter headings and summaries, which facilitate the creation of comprehensive, long-form episodes by segmenting the script into chapters. 
These summaries guide the script generation process, ensuring the episode's overall coherence. 
Concurrently, the \textit{Info Extractor} retrieves the paper's title and the author from the paper. 
The outputs from these assistants are then fed into the \textit{Script Writer}, which produces the dialogue for both the host and the guest speaker based on the chapter plan. 
The completed script is subsequently converted into audio using text-to-speech technology and is mixed with background music to produce the final audio file.

PaperWave uses the OpenAI APIs for the LLM, PDF processing, and TTS.
We used OpenAI's GPT as the LLM~\cite{OpenAI.[n.d.].Models}.
As shown in \autoref{fig:4-2-recording-page}, the model can be selected by the user. 
We implemented the assistants using OpenAI's Assistants API~\cite{OpenAI.[n.d.].AssistantsAPIOverview}.
The assistants are instructed to produce structured output specified by JSON Schema. 
The assistants depend on the File Search~\cite{OpenAI.[n.d.].FileSearch} for PDF content retrieval.
We also used OpenAI's API for TTS~\cite{OpenAI.[n.d.].Textspeech}.

\subsubsection{Web Application}

The web application is the user interface for PaperWave.
The web application is implemented with NextJS~\cite{Vercel.[n.d.].NextjsVercelReactFramework}, a React framework, and NextUI~\cite{NextUI.[n.d.].NextUIBeautifulFastModern}, a UI library for NextJS.
The web application is deployed on Vercel~\cite{Vercel.[n.d.].VercelBuildDeployBest}, and PaperWave CLI for web application is hosted on Google Compute Engine~\cite{Google.2024.ComputeEngineDocumentation} using Docker technology~\cite{Docker.[n.d.].DockerAcceleratedContainerApplication}.
The web application communicates with the backend script via Cloud Firestore~\cite{Google.2024.Firestore}.
When a user requests a \textit{recording}, the options are communicated via Firestore, and the backend script generates the podcast audio file.
The generated podcast is stored in Cloud Storage for Firebase~\cite{Google.2024.CloudStorageFirebase} and can be accessed by the user from the web application.

\end{CJK*}
\end{document}